\documentclass[aps,prd,preprintnumbers,superscriptaddress,nofootinbib,floatfix,twocolumn,notitlepage,reprint]{revtex4-1}
\usepackage[colorlinks=true,citecolor=blue,linkcolor=blue,urlcolor=blue]{hyperref}
\usepackage{graphicx,array}
\usepackage{xcolor}
\usepackage[dvipsnames]{xcolor}
\usepackage{amsmath,amssymb,slashed,latexsym}
\usepackage{bm}
\usepackage{braket}
\usepackage{placeins}
\usepackage{adjustbox}
\usepackage{makecell}
\usepackage[caption=false]{subfig}
\usepackage{cancel}
\usepackage{enumitem}
\usepackage{amsthm}
\usepackage{comment}
\usepackage{multirow} 
\usepackage{tikz}
\usepackage[compat=1.1.0]{tikz-feynman}
\usepackage{soul}

\tikzstyle{block} = [draw, rectangle, 
minimum height=3em, minimum width=6em]

\newcommand*{\rom}[1]{\expandafter\@slowromancap\romannumeral #1@}

\newcommand{\cO}{\mathcal{O}}
\newcommand{\cL}{\mathcal{L}}

\newcommand{\cm}{\mathrm{cm}}

\newcommand{\MeV}{\mathrm{MeV}}
\newcommand{\GeV}{\mathrm{GeV}}
\newcommand{\TeV}{\mathrm{TeV}}

\newcommand{\fm}{\mathrm{fm}}

\def\beq{\begin{equation}}
\def\eeq{\end{equation}}
\def\beqa{\begin{eqnarray}}
\def\eeqa{\end{eqnarray}}

\begin{document}

\preprint{CERN-TH-2026-263}
\title{Probing imperfect axions with nuclear clocks}

\author{Salvatore Bottaro}
\affiliation{Department of Particle Physics and Astrophysics, Weizmann Institute of Science, Rehovot 7610001, Israel}
\affiliation{Center for Theoretical Physics – a Leinweber Institute, Massachusetts Institute of Technology, Cambridge,
MA 02139, USA}

\author{Andrea Caputo}
\affiliation{Theoretical Physics Department, CERN, 1211 Geneva 23, Switzerland}
\affiliation{School of Natural Sciences, Institute for Advanced Study, Princeton, NJ 08540, USA}
\affiliation{Dipartimento di Fisica, ``Sapienza'' Universit\`a di Roma \& Sezione INFN Roma1, Piazzale Aldo Moro
5, 00185, Roma, Italy}

\author{Gilad Perez}
\affiliation{Department of Particle Physics and Astrophysics, Weizmann Institute of Science, Rehovot 7610001, Israel}

\author{Yotam Soreq}
\affiliation{Physics Department, Technion - Israel Institute of Technology, Haifa 3200003, Israel}

\author{Alfredo Stanzione}
\affiliation{Dipartimento di Fisica, ``Sapienza'' Universit\`a di Roma \& Sezione INFN Roma1, Piazzale Aldo Moro
5, 00185, Roma, Italy}

\begin{abstract}
We show that $^{229}$Th nuclear clocks can probe departures from the Peccei--Quinn~(PQ) solution to the strong CP problem. 
Explicit PQ breaking or additional CP-violating interactions can shift the minimum of the axion potential and generate CP-violating couplings to matter, inducing a clock response linear in the axion field. 
We investigate explicit PQ breaking and quark chromo-electric dipole operators as representative realizations of these \textit{imperfect axion} scenarios, consistent with electric dipole moment constraints. 
For axion dark matter, the linear signal oscillates at the Compton frequency and can dominate the quadratic response. 
Even without axion dark matter, the Earth can source a static axion field, enabling fifth-force searches with nuclear clocks. 
\end{abstract}

\maketitle

\section{Introduction}

The Standard Model~(SM) contains two physical CP-odd phases: 
the Cabibbo--Kobayashi--Maskawa phase $\delta_{\rm CKM}$ and the QCD vacuum angle,
\begin{align}
    \label{eq:GGdual}
    \cL 
    \supset 
    \frac{\alpha_s}{8\pi}\bar\theta\, G_{\mu\nu}^a\tilde{G}^{a,\mu\nu}\,,
    \qquad
    \bar{\theta}
    =
    \theta+\arg\det M_q\,
\end{align}
with $G$\,($\tilde G$) the (dual)\,gluon field strength and $M_q$ the quark mass-matrix.
While $\delta_{\rm CKM}$ is order one, neutron electric dipole moment~(EDM) bounds~\cite{Baker:2006ts,Abel:2020pzs}, combined with the lattice result of Ref.~\cite{Liang:2023jfj}, imply $\bar\theta\lesssim10^{-10}$ (see~\cite{Liu:2024kqy,Pospelov:2025vzj,deVries:2026quu}), constituting the strong CP problem. 
Contributions induced within the SM lie far below present sensitivity~\cite{Ellis:1978hq,Georgi:1986kr,Khriplovich:1985jr,Gerard:2012ud}, but this suppression is accidental, resting on the flavour structure and hierarchies of the SM. 
New physics near the TeV scale, as motivated by the hierarchy problem, and by baryogenesis~\cite{Sakharov:1967dj}, generically introduces additional CP violation involving quarks and gluons, which feeds sizably into $\bar\theta$.

The Peccei--Quinn~(PQ) mechanism addresses the problem by promoting $\bar\theta$ to a dynamical field, $\bar\theta\to\bar\theta+a(x)/f_a$, whose QCD potential is minimised at the CP-conserving point~\cite{PhysRevLett.38.1440,Peccei:1977ur,Weinberg:1977ma,PhysRevLett.40.279}.
The above mechanism, however, becomes imperfect for two main reasons. 
First, the presence of explicit PQ-violating~(PQV) terms, as motivated by the axion quality problem~\cite{Kamionkowski:1992mf,Holman:1992us,PhysRevD.46.539}, can misalign the axion potential and leave a residual angle $\theta_{\rm eff}\equiv\bar\theta+\langle a\rangle/f_a\neq0$.
Furthermore, new CP-violating interactions involving non-decoupling SM operators may shift the axion vacuum~\cite{Shifman:1979if,Pospelov:1997uv,Pospelov:1999mv,Pospelov:2000bw}.

The implications of axion imperfections have profound and potentially dramatic impact on clock-based searches of the axion that have not been studied systematically thus far. 
In particular, as we demonstrate below, there are several new effects that can be probed using nuclear clocks.
The low energy isomeric transition of $^{229}$Th is highly sensitive to time variations of hadronic parameters~\cite{Flambaum:2006yx}, which in the CP-conserving limit was shown to be sensitive to the second power of the oscillating axion field~\cite{Kim:2022ype,Kim:2023pvt}. 
Imperfect axions lead to two new types of signal in clocks through linear power in the axion dark matter~(DM) field, and via inducing long range forces.

As explained below, the overall clock response to imperfect axions considered in this work can be written as
\begin{align}
    \label{eq:deltanunu_general}
    \frac{\delta\nu_{\rm Th}}{\nu_{\rm Th}}
    =
    K \left( 2\, x^*  \,\theta_{\rm eff}\frac{\delta a}{f_a}
    + x_{\pi} \frac{\delta(a^2)}{f_a^2} \right)\,,
\end{align}
where $K\sim10^{4}$--$10^5$ is the enhancement factor of the thorium transition~\cite{Caputo:2024doz,Beeks:2024xnc},  
$x_\pi\simeq7\times10^{-3}$~\cite{Kim:2022ype} is set by the nucleon mass response to the pion mass variation due to the axion.
For PQ breaking alone, the hadronic parameters depend on the axion only through $\Theta\equiv \theta_{\rm eff}+a/f_a$, so that the response is a function of $\delta(\Theta^2)$ with $x^*=x_\pi$. 
Generic CP-violating SM operators contribute to $x^*$ both through the shift of the axion minimum and through a direct modification of the axion dependence of the hadronic parameters, so that in general $x^*\neq x_\pi$. However, for the cases below $x^*\sim\cO(x_\pi)$, see discussion.

Nuclear clocks, or laser excitation of $^{229}$Th~\cite{Fuchs:2024xvc}, therefore probe a combination of CP-violating Wilson coefficients that is correlated with, but not equivalent to, the one measured by EDMs, giving rise to a new signal that is potentially visible already at current sensitivity.
To demonstrate it, we note that the linear and quadratic signals differ in two respects. 
For axion DM, the linear term in Eq.~\eqref{eq:deltanunu_general} oscillates at the Compton frequency $m_a$, rather than at $2m_a$ as the quadratic term, and, since its amplitude scales as $1/f_a$ rather than $1/f_a^2$, at fixed $m_a$ its reach improves with the exposure time as $f_a\propto t^{1/2}$ rather than $t^{1/4}$. 
As a result, the linear contribution overtakes the quadratic one after a total exposure time
\begin{align}
    \label{eq:tcross}
    t_{\text{cross}} 
    \simeq 
    80 \,\text{days} 
    &
    \left( \frac{7 \times 10^2}{K\, x^*}  \right)^{4}\left( \frac{ K\, x_\pi  }{7 \times 10^{2}} \right)^{2}
    \\ \nonumber
    &
    \left( \frac{\delta \nu/\nu}{5 \times 10^{-14}} \right)^{2}
    \left( \frac{10^{-9}}{\theta_{\rm eff}} \right)^{4}
    \,,
\end{align}
using the 23-hour data taking sensitivity of  Ref.~\cite{DeCol:2026hnl} (for masses lower than about the inverse of a day see the bound of~\cite{Arakawa:2026mls}).

Furthermore, the linear term in Eq.~\eqref{eq:deltanunu_general} and the scalar axion--nucleon coupling that mediates long range interactions~\cite{Moody:1984ba} have the same origin. Independently of whether the axion is DM, the axion field sourced by ordinary matter therefore shifts the clock frequency through the same term, so that nuclear clocks can also be used as fifth-force probes~\cite{Delaunay:2025lgk}, covering uncharted territory of the imperfect axion parameter space.

\section{Imperfect axions}
\label{sec:imperfect_axions}

We consider two classes of imperfect axions. 
In the first, explicit PQ-violating interactions displace the axion vacuum from the CP-conserving point, and the clock response is fixed by the resulting total angle $\Theta$ alone. 
We study this case both in isolation and together with quark electric dipole moments~(EDMs), which relax the hadronic EDM bound on $\theta_{\rm eff}$ without affecting the hadronic response to the axion. 
In the second class, represented here by quark chromo-electric dipole moments~(CEDMs), the operators that displace the vacuum further directly modify the axion dependence of hadronic parameters.

We begin with the first class. 
The PQ symmetry is explicitly broken by higher-dimensional operators built solely from the PQ field, $\Phi=(f_a/\sqrt{2})e^{ia/f_a}$. 
These operators of the form $e^{i\delta}\Phi^n/\Lambda^{n-4}+{\rm h.c.}$, with $\delta\sim 1$, are motivated, for instance, by the expectation that quantum gravity admits no exact global symmetries~\cite{Witten:2017hdv,Banks:2010zn,Harlow:2018tng,Abbott:1989jw,Coleman:1989zu,Giddings:1987cg,Lee:1988ge}. 
These result in axion-potential terms which are misaligned with the QCD one, shifting the axion's VEV away from the CP-conserving value by 
\begin{align}
    \label{eq:theta_PQV}
    \theta_{\rm eff}^{\cancel{\rm PQ}}
    \sim
    2n
    \left(
    \frac{f_a}{\sqrt{2}\Lambda}
    \right)^n
    \frac{\Lambda^4}{\chi_{\rm QCD}} ,
\end{align}
where $\chi_{\rm QCD}$ is the QCD topological susceptibility~\cite{Shifman:1979if,GrillidiCortona:2015jxo}.
For $\Lambda\simeq M_{\rm Pl}$, EDM bounds on this residual angle require all such operators up to dimension $n\sim 10$ or higher, depending on $f_a$, to be absent, which is known as the axion quality problem~\cite{Kamionkowski:1992mf,Holman:1992us,PhysRevD.46.539}.

Within this first class, we consider two possibilities. 
Benchmark~(\textit{i}), which we refer to as \textit{PQ-breaking}, is the minimal case in which explicit PQ breaking is the only relevant new source of CP violation, so that EDM constraints directly bound the residual angle in Eq.~\eqref{eq:theta_PQV}.
Considering the neutron and Hg EDM bounds~\cite{Pendlebury:2015lrz,Abel:2020pzs,Graner:2016ses} and profiling over the lattice uncertainties~\cite{Liu:2024kqy} (details in Appendix~\ref{app:benchmarks}) leads to 
\begin{align}
    |\theta_{\rm eff}^{\cancel{\rm PQ}}|
    \lesssim 
    1.5\times10^{-10} \, .
\end{align}
Benchmark~(\textit{ii}), denoted \textit{PQ-breaking+EDM}, includes additional quark EDM operators,
\begin{align}
    \label{eq:EDM}
    \cL_{\rm EDM}
    \supset
    -\frac{i}{2} \sum_{q=u,d,s}
    d_q\,\bar q \sigma^{\mu\nu}\gamma_5 q\,F_{\mu\nu},
\end{align}
which contribute independently to hadronic EDMs. 
Their presence can therefore allow cancellations in EDM observables and substantially weaken the bound on the PQ-breaking contribution to $\theta_{\rm eff}$, such that 
\begin{align}
    |\theta_{\rm eff}^{\cancel{\rm PQ}}|
    =
    10^{-9} \, ,
\end{align}
is allowed by the above constraints. 
In the setup considered here, however, they do not modify the leading axion dependence of the hadronic parameters relevant for the clock response. 
Thus, in both cases the leading axion dependence is controlled by the shifted total angle $\Theta$ with $\theta_{\rm eff}\to \theta_{\rm eff}^{\cancel{\rm PQ}}$.

Benchmark~(\textit{iii}), which we call \textit{CEDM}, represents the second class 
in which the vacuum displacement does
not by itself determine the leading axion response. 
We illustrate this possibility with quark CEDM operators,
\begin{align}
    \label{eq:CEDM}
    \cL_{\rm CEDM}
    \supset
    -\frac{i g_s}{2} \sum_{q=u,d,s}
    \widetilde d_q\,
    \bar q_L\sigma^{\mu\nu}G^a_{\mu\nu}T^a q_R
    +{\rm h.c.} \, .
\end{align}
These operators arise naturally in extensions of the SM with new CP-violating phases, such as supersymmetric models (see e.g.~\cite{Nakai:2016atk,Kaneta:2023wrl} and references therein) and warped-extra-dimension/composite Higgs models~\cite{Agashe:2004cp,Delaunay:2012cz,Konig:2014iqa}.

Although they preserve the PQ symmetry, CEDMs modify the QCD vacuum energy and therefore displace the axion minimum. 
The resulting residual angle can be computed either in chiral perturbation theory, through a consistent vacuum alignment~\cite{Dekens:2022gha}, or with functional methods, from the QCD vacuum energy in the presence of CP-odd sources~\cite{Shifman:1979if,Pospelov:1997uv,Pospelov:1999mv,Pospelov:2000bw,Okawa:2021fto}.
To leading order in the CEDM coefficients, one finds
\begin{align}
    \label{eq:theta_CEDM}
    \theta_{\rm eff}^{\widetilde d}
    =
    \frac{m_0^2}{2}
    \sum_q\frac{\widetilde d_q}{m_q},
\end{align}
where $m_0^2\simeq0.8\,\GeV^2$ parametrizes the mixed quark--gluon condensate~\cite{Pospelov:2005pr,Belyaev:1982sa,Belyaev:1984ic}.
After imposing the experimental constraints on the Hg and neutron EDMs, we find the maximal allowed value
\begin{align}
    |\theta_{\rm eff}^{\widetilde d}|
    \simeq 
    8.8\times10^{-10} \, ,
\end{align}
with a corresponding new-physics scale of $\Lambda=\sqrt{m_q/|\widetilde d_q|}\sim\cO(20\,\TeV)$, see details in Appendix~\ref{app:benchmarks}.

\section{Implications for nuclear clock experiments}
\label{sec:nuclear_clock}

We parametrize the response of the thorium clock to a variation of the nucleon mass as $\delta\nu_{\rm Th}/\nu_{\rm Th}=K\,\delta m_N/m_N$. 
Expanding $m_N$ to second order in the axion field yields, up to an irrelevant overall sign, the two terms of Eq.~\eqref{eq:deltanunu_general}, whose coefficients we now discuss\footnote{The linear coupling of a CP-violating QCD axion to clocks was considered in Ref.~\cite{Arvanitaki:2014faa}, where the axion is treated as a dilaton, so that the hadronic parameters follow the quark masses through their sigma terms. This treatment does not explicitly include the direct contributions of the CP-violating operators to the axion dependence of the hadronic masses, or the term quadratic in the axion field. Ref.~\cite{Plakkot:2023pui}, on the other hand, considers signals from the time variation of $\alpha_{\rm em}$, $m_e$ and the nuclear magnetic moment, so that the hadronic sensitivity of the $^{229}$Th transition, on which this work relies, is not exploited.}. 
The quadratic term in Eq.~\eqref{eq:deltanunu_general} is the CP-even response of Ref.~\cite{Kim:2022ype}. 
The pion mass decreases as $\delta\ln m_\pi^2=-\tfrac{z}{2}(a/f_a)^2$, with $z\equiv m_um_d/(m_u+m_d)^2$, and the nucleon mass follows it through the sigma term, $\delta m_N=(\sigma_{\pi N}/m_\pi^2)\,\delta m_\pi^2$, where $\sigma_{\pi N}=-4c_1B(m_u+m_d)=59\,\MeV$ \cite{Hoferichter:2015dsa}, so that $x_\pi=z\,\sigma_{\pi N}/2m_N\simeq 7\times10^{-3}$, independently of the CP-odd sources.

The linear term requires the first derivative of the nucleon mass with respect to the axion field. 
Treating explicit PQ breaking and the CEDMs together (see Appendix~\ref{app:chiPT}), it reads
\begin{align}
    \label{eq:nucleon_mass_derivative}
    \left.\frac{\partial m_N}{\partial(a/f_a)}\right|_{\rm vac}
    \!\!\!
    &=
    4Bc_1m_*\,\theta_{\rm eff}^{\cancel{\rm PQ}}
    +4B(c_1 - \tilde c_1)m_*\theta_{\rm eff}^{\tilde d} 
    \nonumber\\
    &=
    -z\,\sigma_{\pi N}\left(\theta_{\rm eff}^{\cancel{\rm PQ}}+\theta_{\rm eff}^{\tilde d}\right)
    -4Bm_*\,\tilde c_1\,\theta_{\rm eff}^{\tilde d},
\end{align}
where $m_*=m_um_d/(m_u+m_d)$. 
The first term is the indirect contribution: 
the tilted vacuum shifts the quark-mass term as a $\theta$ angle of size
$\theta_{\rm eff}^{\cancel{\rm PQ}}+\theta_{\rm eff}^{\tilde d}$ would, and the nucleon mass follows the corresponding pion-mass variation through the sigma term. 
The second is the direct contribution of the CEDM operator, controlled by the independent Low Energy Constant~(LEC) $\tilde c_1$, see Appendix~\ref{app:chiPT} for details.

The benchmarks (\textit{i})--(\textit{ii}) correspond to taking the limit $\theta_{\rm eff}^{\tilde d}\to0$, such that $x^{*}=x_{\pi}$ in Eq.~\eqref{eq:deltanunu_general}, while benchmark (\textit{iii}) corresponds to the limit $\theta_{\rm eff}^{\rm \cancel{\rm PQ}}\to 0$ and $x^*=(1-\tilde c_1/c_1)\,x_\pi$. 
There is no established lattice determination for the constant $\tilde c_1$ (see Refs.~\cite{Khatsimovsky:1987bb,Jin:1993up,Pospelov:2001ys,Seng:2018wwp} for alternative attempts).  
Thus, for simplicity, we assume $|1-\tilde c_1/c_1|=\cO(1)$ and adopt $|x^*|\simeq x_\pi$ for benchmark~(\textit{iii}), while keeping in mind that $\cO(1)$ corrections might enhance or reduce the total effect.

Importantly, we stress that Eq.~\eqref{eq:nucleon_mass_derivative} is specific to the concrete benchmarks considered in this work. 
In the presence of multiple CP-violating operators, such as CEDMs and CP-odd four-quark interactions, the linear combination of Wilson coefficients appearing in the term linear in the axion field in the nucleon mass would not be proportional to the combination inducing $\theta_{\rm eff}$~\cite{Bertolini:2020hjc,DiLuzio:2024ctr}. 
A systematic exploration of this more general CP-violating parameter space is left for future work.

We now consider the time-dependent clock signal generated when the axion constitutes all of the dark matter~\cite{Preskill:1982cy,Abbott:1982af,Dine:1982ah}. 
For sufficiently light axions, the large occupation number allows the field to be described as a classical wave oscillating about its vacuum, $a(t)=a_0\cos(m_at+\phi)$, with amplitude $a_0=\sqrt{2\rho_{\rm DM}}/m_a$ fixed by the local dark matter
density, $\rho_{\rm DM}=0.4\,{\GeV\,\cm^{-3}}$~\cite{Benito:2020lgu}. 
For $\theta_{\rm eff}=0$, the leading time-dependent signal is then quadratic
in the axion field, oscillates at $2m_a$, and scales as $f_a^{-2}$ at fixed $m_a$. A nonvanishing residual angle generates instead the linear term, which oscillates at $m_a$ and scales as $f_a^{-1}$.
The latter gives the expected sensitivity 
\begin{align}
    \label{eq:fainvDM}
\frac{1}{f_a}\lesssim &\left(\frac{2.8\times10^{-17}}{\rm GeV}\right)\,\left(
\frac{\delta\nu_{\rm Th}/\nu_{\rm Th}}{10^{-19}}\right)\, \nonumber\\
&\qquad\quad \left(\frac{m_a}{10^{-15}\,{\rm eV}}\right)\,
\left(\frac{7\times10^2}{K\,x^*}\right)\,\left(
\frac{10^{-9}}{\theta_{\rm eff}}\right)\,.
\end{align}
At large masses, this sensitivity deteriorates because of the finite dark-matter coherence time and the averaging of rapid oscillations over a measurement cycle~\cite{Banerjee:2020kww}.
At small masses, the loss of sensitivity arises because the finite observation time is insufficient to resolve a complete oscillation~\cite{Barontini:2021mvu}. 

\begin{figure*}[htp!]
    \centering
    \includegraphics[width=0.47\linewidth]{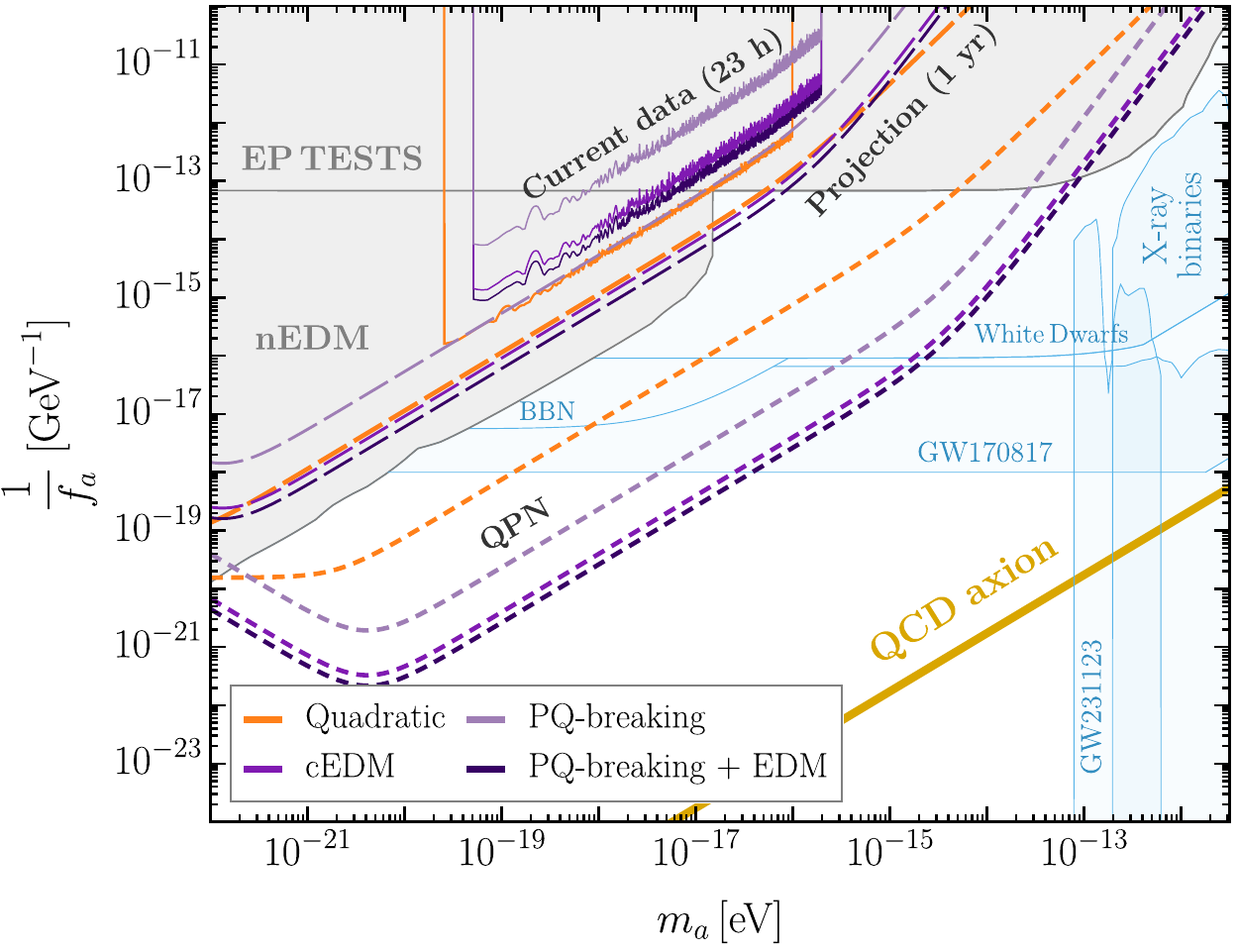}\qquad
    \includegraphics[width=0.47\linewidth]{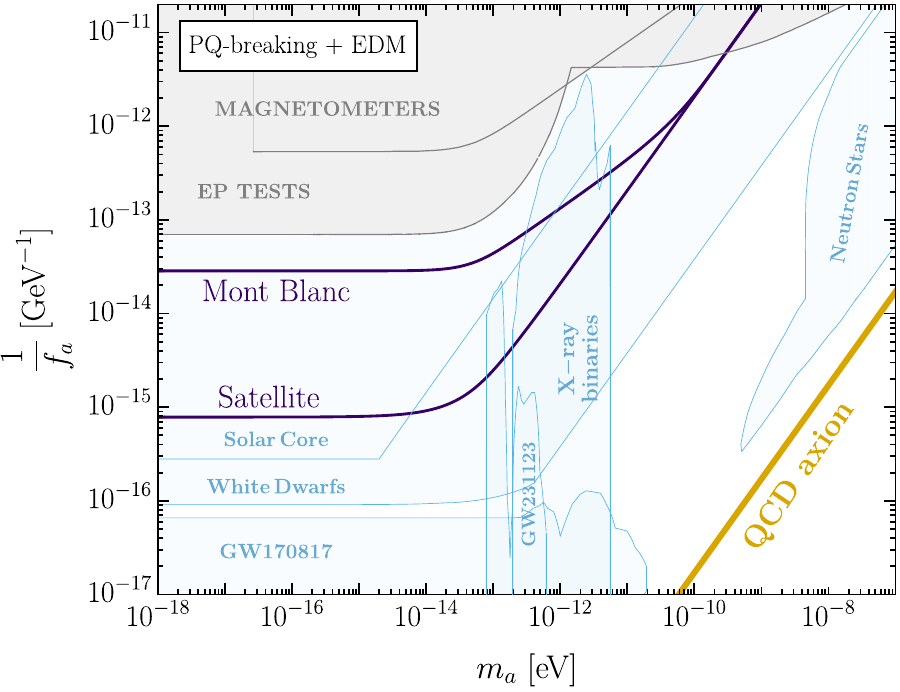}
    \caption{Constraints and projected sensitivities in the $m_a$--$f_a^{-1}$ plane. (\textit{Left}) The orange and purple curves correspond to the quadratic and linear terms in Eq.~\eqref{eq:deltanunu_general}. Different purple curves are obtained for the benchmarks (\textit{i})--(\textit{iii}) discussed in the main text, with the last two cases slightly offset for readability. (\textit{Right}) Projected sensitivities from fifth force searches using nuclear clocks for imperfect axions in the \textit{PQ-breaking+EDM} scenario. The solid purple lines show the sensitivity obtained by placing one clock on the surface of Earth, and a second one on a satellite in space or at the top of Mont Blanc. The diagonal golden line denotes the standard QCD-axion relation. Gray lines show constraints from searches for an oscillating neutron EDM~\cite{Abel:2017rtm}, Equivalence Principle~(EP) tests (assuming the largest $\theta_{\rm eff}$ in the \textit{left} panel)~\cite{Berge:2017ovy,Smith:1999cr,Kapner:2006si,Lee:2020zjt} and magnetometer experiments~\cite{Zhang:2023qmu,Bulatowicz:2013hf,Petukhov:2010dn,Tullney:2013wqa,Lee:2018vaq}. Light blue regions show astrophysical bounds from White Dwarfs \cite{Balkin:2022qer}, Neutron Stars \cite{Gomez-Banon:2024oux,Kumamoto:2024wjd}, Solar Core \cite{Hook:2017psm,DiLuzio:2021gos,DiLuzio:2021pxd}, and superradiance searches \cite{Caputo:2025oap,Witte:2024drg} (see also \cite{Ning:2026ebu,Khalaf:2026ovs}). }
    \label{fig:clocks}
\end{figure*}

The results are displayed in the left panel of Figure~\ref{fig:clocks}, where we show the existing constraints and the projected sensitivity of the $^{229}$Th nuclear clock in the $m_a$--$f_a^{-1}$ plane, for $K=10^5$. 
The orange curves correspond to the limit $\theta_{\rm eff}\to0$, while the purple curves show the reach associated with the linear term for the benchmarks (\textit{i})--(\textit{iii}), with the curves corresponding to the last two scenarios slightly offset ($\sim 10\%$) for readability. 
The solid curves show the sensitivity achieved in the recent experimental analysis of Ref.~\cite{DeCol:2026hnl}, corresponding to $\delta\nu_{\rm Th}/\nu_{\rm Th}\simeq5\times10^{-14}$ with 23 hours of data (see also Ref.~\cite{Huang2026NuclearClock} for a second, independent $^{229}$Th nuclear clock).
The long-dashed curves show the projected sensitivity for the same experimental setup after one year of exposure, assuming
white-noise scaling, while the short-dashed curves denote the quantum-projection-noise~(QPN) reach~\cite{Kim:2023pvt,Kim:2022ype,Zhang:2024ngu}, $\delta\nu_{\rm Th}/\nu_{\rm Th}\simeq10^{-19}$, for one day of data taking.
We also show astrophysical constraints coming from the sourcing in White Dwarfs~\cite{Balkin:2022qer}, Neutron Stars \cite{Gomez-Banon:2024oux,Kumamoto:2024wjd}, Solar Core \cite{Hook:2017psm,DiLuzio:2021gos,DiLuzio:2021pxd}, and superradiance searches ~\cite{Caputo:2025oap,Witte:2024drg,Ning:2026ebu,Khalaf:2026ovs}. In the case of non-negligible axion-photon coupling (which is typically expected for untuned QCD axion models), one also has strong bounds from the death line of rotation-powered pulsars~\cite{Witte:2025ilt}.

In the figures we compare nuclear-clock sensitivities with terrestrial, astrophysical, and cosmological constraints. 
It is important to note however that the latter typically involve further theoretical inputs, distinct systematic uncertainties, and may probe different combinations of axion couplings. 
For instance, unlike the clocks, superradiance bounds depend sensitively on axion self-interactions, BBN constraints on the cosmological evolution of the axion field at early times, White-Dwarf bounds may be affected by axion–electron couplings, and gravitational-wave constraints by uncertainties in the neutron-star equation of state. 
Thus, while complementary, these bounds are not always directly comparable to nuclear-clock searches, which provide a controlled laboratory probe of specific hadronic couplings.

Crucially, nuclear clocks can probe imperfect axions even if they do not constitute dark matter, following the proposal of Ref.~\cite{Delaunay:2025lgk} to search for fifth forces with nuclear clocks. 
In the presence of an imperfect axion, a static macroscopic distribution of matter acts as a source for the axion field. 
Approximating the Earth as a uniform-density sphere of radius $R_\oplus$ and mass $M_\oplus$, the solution of the static Klein--Gordon equation outside the Earth is~\cite{Delaunay:2025lgk}
\begin{align}
    \label{eq:earth_profile}
    a_\oplus(r)
    =
    2\,\theta_{\rm eff}\, x_\pi \frac{M_\oplus}{f_a}
    \frac{e^{-m_ar}}{4\pi r}
    F(m_aR_\oplus),
\end{align}
where $F(x)=3(x\cosh x-\sinh x)/x^3$.
The source is the same derivative of the nucleon mass that controls the linear clock response, Eq.~\eqref{eq:nucleon_mass_derivative}, and here we fix $x^*= x_\pi$.
A comparison of the clock frequency at two distances $r_1$ and $r_2$ from the center of the Earth then measures $a_\oplus(r_1)-a_\oplus(r_2)$. 
For $r_1=R_\oplus$, $a(r_2)\ll a(R_\oplus)$  and $m_aR_\oplus\ll1$, combining Eqs.~\eqref{eq:deltanunu_general} and \eqref{eq:earth_profile} gives the sensitivity
\begin{align}
    \frac{1}{f_a}
    \lesssim\,
    &\frac{7.9\times10^{-16}}{\rm GeV}\,
    \left( \frac{10^{-9}}{\theta_{\rm eff}}\right)
    \nonumber \\
    & \left(\frac{\delta\nu_{\rm Th}/\nu_{\rm Th}}{10^{-19}}\,
    \frac{10^{5}}{K}\right)^{1/2}
    \left(
    \frac{7\times10^{-3}}{x_\pi}\right)\,,
\end{align}
where $R_\oplus=6371\,\rm km$, $M_\oplus=5.97 \times 10^{24}\,\rm kg$~\cite{NASAEarth}.

In Figure~\ref{fig:clocks} right, we show the
projected nuclear clock sensitivities on $f_a^{-1}$ from fifth-force searches assuming the \textit{PQ-breaking+EDM} scenario (see Appendix \ref{sec:app_scalings}  for details on the other benchmarks). The projected bounds are derived by assuming the QPN sensitivity and  two different locations for the far-away clock: on a satellite in space and at the top of Mont Blanc. In both cases, nuclear clock bounds are expected to overcome current constraints from Equivalence Principle (EP) tests \cite{Berge:2017ovy,Smith:1999cr,Kapner:2006si,Lee:2020zjt} and magnetometer experiments \cite{Zhang:2023qmu,Bulatowicz:2013hf,Petukhov:2010dn,Tullney:2013wqa,Lee:2018vaq}. 

\section{Outlook}

We have shown that CP violation qualitatively changes the physics accessible to nuclear clocks: imperfect axions generate a response linear in the axion field, enhancing searches for axion dark matter and enabling matter-sourced fifth-force signals.  

Throughout this work, we have expressed the clock response through $\delta m_N/m_N$. 
The $^{229}$Th transition is, however, expected to be sensitive to the variation of several hadronic scales such as the nucleon, $\Delta$, pion and vector-meson masses, see e.g. Ref.~\cite{Kim:2022ype}. 
For the quadratic axion dependence, all of them are tied to $\delta m_\pi^2$. 
Additional CP-odd sources can generate linear responses that are not captured by this common dependence: besides the vacuum-realignment piece, each hadron mass receives a direct contribution governed by matrix elements of the relevant higher-dimensional operators, so that the variations of $m_N$, $m_\Delta$, $m_\pi$ and $m_V$ become independent. 
The chromo-EDM is the striking example: the pion mass has no term linear in the axion field at leading chiral order, whereas the nucleon mass does.
A complete treatment of the clock response to imperfect axions would thus require additional hadronic matrix elements beyond the nucleon sigma term, and the resulting sensitivity coefficient may differ from the CP-even one by an $\cO(1)$ factor, possibly including its sign. 
We leave this investigation for future work.

An important next step is to extend our analysis to the complete set of CP-violating operators, in particular CP-odd four-fermion interactions.
This will require improved knowledge of the relevant hadronic matrix elements, many of which remain poorly determined, and will enable a systematic study of correlations with EDMs and spin-dependent CP-violating forces.
Most intriguingly, dimension-six CP-violating interactions can arise from new physics probed at the HL-LHC and future colliders.  
Nuclear clocks therefore offer a unique bridge between axion and fifth-force searches, EDMs, and the energy frontier, opening the possibility of correlated signals from TeV-scale collisions to the $^{229}$Th nuclear transition.

\section*{ACKNOWLEDGMENTS}
AS would like to thank Martin A. Mojahed for useful conversations. SB is grateful to Konstantin Springmann for clarifications on the astrophysical bounds. AC and AS are supported by an ERC STG grant (“AstroDarkLS”, grant No. 101117510). AC acknowledges the hospitality of the Institute for Advanced Study supported by an anonymous donor during the final stages of this project. AC and AS also acknowledge the Weizmann Institute of Science for hospitality and support during the initial phases.
GP is supported by the Israel Science Foundation (ISF), Minerva, the NSF-BSF, and the European Research Council (ERC, DM-Dawn, Grant Agreement No. 101199868).
YS is supported by the ISF (grant No. 597/24) and by BSF (grant No. 2024091). 
The authors acknowledge the use of Claude (Anthropic) and ChatGPT (OpenAI) for cross-checks, language editing, and figure code.

\newpage

\appendix

\section*{Appendices}

\section{Fifth-force searches with nuclear clocks in other axion benchmarks}
\label{sec:app_scalings}

\begin{figure*}[htp!]
    \centering
    \includegraphics[width=0.47\linewidth]{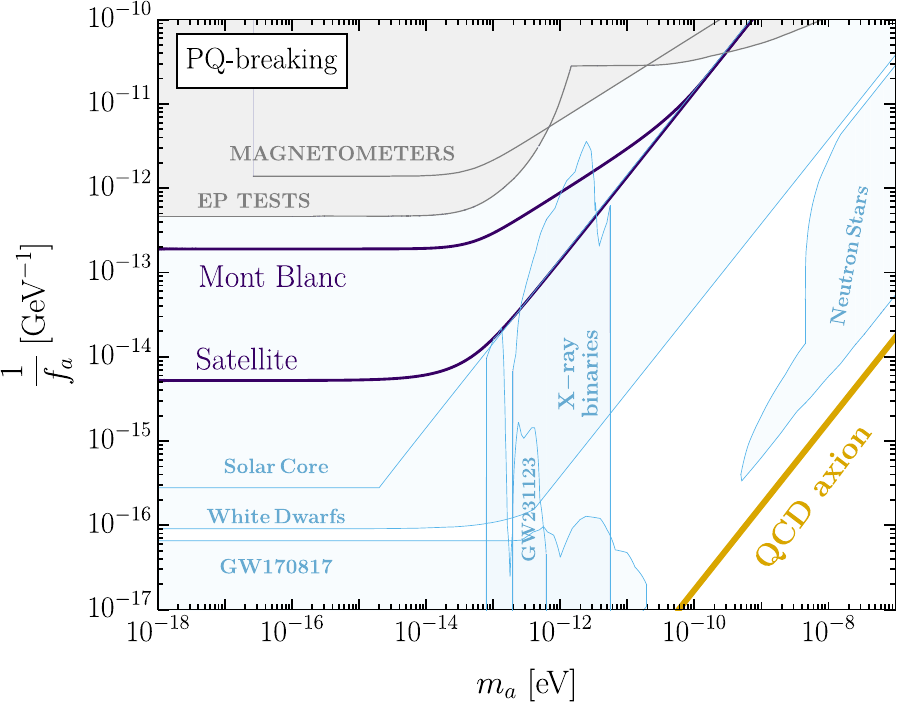}\qquad\includegraphics[width=0.47\linewidth]{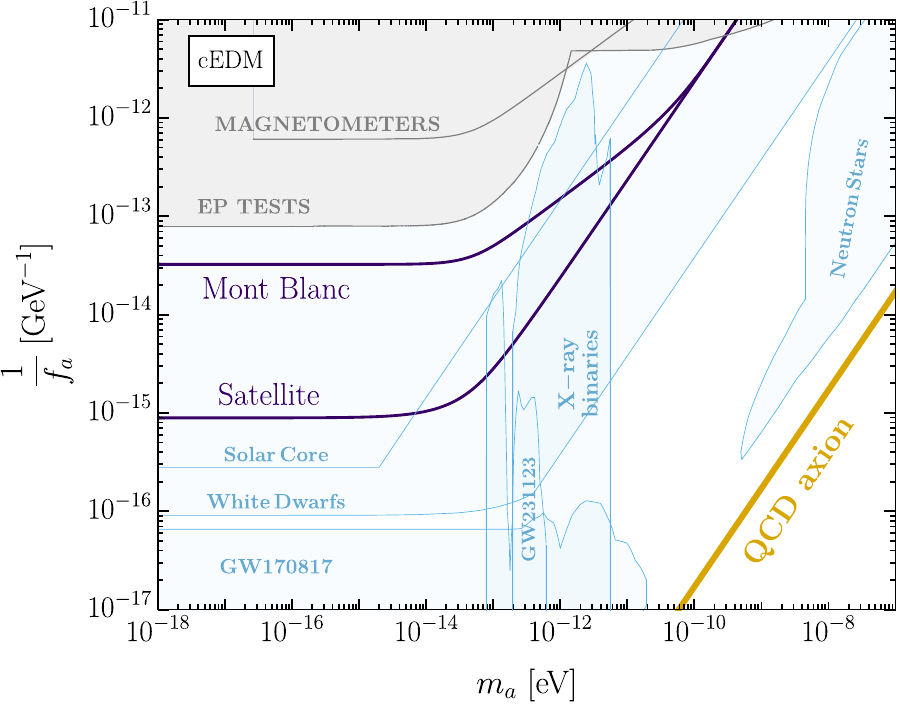}
    \caption{Constraints and projected sensitivities in the $m_a$--$f_a^{-1}$ plane from fifth force searches using nuclear clocks for imperfect axions in the \textit{PQ-breaking} (\textit{left}) and \textit{cEDM} (\textit{right}) scenarios. Bounds and their labels are defined in Figure \ref{fig:clocks}. }
    \label{fig:clocks_app}
\end{figure*}

In this appendix, we briefly discuss how the bounds and sensitivities on $f_a^{-1}$ as a function of $m_a$ change with $\theta_{\rm eff}$. All the bounds from astrophysical probes (White Dwarfs, Neutron Stars, Solar Core, Superradiance and inspiralling binaries) are derived from the CP-even axion potential and the axion–nucleon interactions
\begin{align}
    \mathcal{L}
    \supset 
    -\sigma_{\pi N}\bar{N}N\left(\sqrt{1-4z\sin^2\left(\frac{a}{2f_a}\right)}-1\right) \, ,
\end{align}
and are thus independent of $\theta_{\rm eff}$. 
EP tests, on the other hand, probe the CP-violating scalar axion-nucleon coupling $g_{aN}^Sa\bar{N}N$, where $g_{aN}^S=(\theta_{\rm eff}/f_a)z \sigma_{\pi N}$~\cite{DiLuzio:2023lmd}. 
Therefore, we use this simple $f_a^{-1}\propto1/\theta_{\rm eff}$ scaling to translate the EP tests bounds in the $f_a^{-1}$-$m_a$ plane. 
Notice that in Figure~\ref{fig:clocks} \textit{left}, we directly showed the most conservative bound among the three benchmark scenarios. 
Finally, magnetometer-based experiments exploit monopole-dipole interactions among nuclei mediated by axions proportional to $g_{aN}^Sg_{aN}^P$, where $g_{aN}^P=C_nm_N/f_a$ is the pseudoscalar axion-nucleon coupling $g_{aN}^Pa\bar{N}i\gamma_5N$. 
The coefficient $C_n=\cO(1)$ in typical axion models and is taken to be unity in this work. 
Therefore, magnetometer bounds will all scale as $f_a^{-1}\propto1/\sqrt{\theta_{\rm eff}}$. 
Analogously to Figure~\ref{fig:clocks} \textit{right} for benchmark~(\textit{ii}), in Figure \ref{fig:clocks_app}, we show the comparison between projected sensitivities from fifth-force searches and current astrophysical, EP tests and magnetometer bounds on $f_a^{-1}$  also for benchmarks~(\textit{i}) and (\textit{iii}).

\section{Baryon-$\chi_{\rm PT}$, induced $\theta_{\rm eff}$, and the scalar\\ axion--nucleon coupling}
\label{app:chiPT}

We match the quark-mass and QCD-angle of Eq.~\eqref{eq:GGdual}, together with the CEDM operators of Eq.~\eqref{eq:CEDM}, onto the two-flavor $\chi_{\rm PT}$ Lagrangian augmented with an axion. 
Explicit PQ breaking is described by an additional potential $V_{\cancel{\rm PQ}}(a)$. 
The relevant nonderivative interactions
are~\cite{Dekens:2022gha}
\begin{align}
    \label{eq:chiral_mass_CEDM}
    \cL_\pi^{(2)}
    =&
    \frac{F_0^2}{4}\mathrm{Tr}
    \left[(\chi_\Theta+\tilde\chi_\Theta)U^\dagger
    +\mathrm{h.c.}\right],
    \nonumber\\
    \cL_N^{(2)}
    =&
    \left(c_1\mathrm{Tr}\chi_{\Theta,+}
    +\tilde c_1\mathrm{Tr}\tilde\chi_{\Theta,+}\right)\bar NN
    \nonumber\\
    &+\bar N\left(
    c_5\widehat{\chi}_{\Theta,+}
    +\tilde c_5\widehat{\tilde\chi}_{\Theta,+}
    \right)N \, ,
\end{align}
where $U=u^2=\exp(i\boldsymbol{\pi}\cdot\boldsymbol{\tau}/F_0)$, $X_+=u^\dagger Xu^\dagger+uX^\dagger u$, and $\widehat X=X-\tfrac12\mathrm{Tr}(X)\,\mathbf{1}$. 
An anomalous axial rotation allows one to introduce the total QCD angle through the spurions $\chi_\Theta=2B\mathcal M e^{iQ\Theta}$, $\tilde\chi_\Theta=-2i\tilde B\widetilde{\mathcal D}e^{iQ\Theta},$ with $\mathcal M=\mathrm{diag}(m_u,m_d)$,
$\widetilde{\mathcal D}=\mathrm{diag}(\tilde d_u,\tilde d_d)$, and $Q=\mathrm{diag}(c_u,c_d)$, where $c_q=m_*/m_q$ and $m_*=m_um_d/(m_u+m_d)$.
We use $r\equiv\tilde B/B=m_0^2/2\simeq0.4\,\GeV^2$\cite{Belyaev:1982sa}. 
The baryonic coefficients $c_{1,5}$ and $\tilde c_{1,5}$ are independent low-energy constants, fixed by the nucleon matrix elements of the quark-mass operator and the chromomagnetic counterparts of the CEDM operator.

These interactions also generate the dimensionless
CP-odd pion--nucleon couplings~\cite{Dekens:2022gha},
\begin{align}
    \label{eq:CPV_pion_nucleon}
    \cL_{\pi N}^{\rm CPV}
    =
    \bar g_0\,\bar N\boldsymbol{\tau}\cdot\boldsymbol{\pi}N
    +\bar g_1\,\pi^0\bar NN \, ,
\end{align}
that will be relevant for the discussion of the hadronic EDM bounds in Appendix~\ref{app:benchmarks}.

In order to determine the vacuum, we substitute $\langle U\rangle=\mathrm{diag}(e^{i\varphi_u},e^{i\varphi_d})$, with $\varphi_u+\varphi_d=0$, and we define $x_q=\varphi_q-c_q\Theta$, so that $x_u+x_d=-\Theta$. The potential thus reads
\begin{align}
    V
    =
    -F_0^2B\!\!\sum_{q=u,d}\!\!
    \left[m_q\cos x_q-r\tilde d_q\sin x_q\right]
    \!+\!V_{\cancel{\rm PQ}}(a) \, .
\end{align}
Minimizing with respect to the neutral pion and axion gives $x_q^{(0)}=-r\tilde d_q/m_q-c_q\theta_{\rm eff}^{\rm PQ}$, where $\theta_{\rm eff}^{\cancel{\rm PQ}}$ is the displacement of the axion minimum induced by $V_{\rm PQ}$, i.e.\ the residual angle of the main text for $\tilde d_q=0$. 
Hence,  the total residual angle is $\theta_{\rm eff}\equiv-\sum_q x_q^{(0)}
=\theta_{\rm eff}^{\tilde d}+\theta_{\rm eff}^{\cancel{\rm PQ}}$, where $\theta_{\rm eff}^{\tilde d}$ matches Eq.~\eqref{eq:theta_CEDM}. 

Expanding around this point, the isoscalar nucleon mass, $m_N=(m_p+m_n)/2$, depends separately on the two baryonic LECs,
\begin{align}
    m_N
    =
    m_N^{(0)}\!-\!4B\sum_q
    \!\left[c_1m_q\cos x_q
    -r\tilde c_1\tilde d_q\sin x_q\right]  ,
\end{align}
and its derivative is given in Eq.~\eqref{eq:nucleon_mass_derivative}, see also discussion in the main text.

\section{Numerical benchmarks}
\label{app:benchmarks}

We first assess the EDM constraints on a residual angle generated by explicit PQ breaking, both in isolation and in the presence of an additional quark EDM, corresponding to the scenarios (\textit{i}) and (\textit{ii}) discussed in the main text. 
The quark EDMs, see Eq.~\eqref{eq:EDM}, contribute to the nucleon EDMs as
\begin{align}
    d_N
    =
    \theta_N\theta_{\rm eff}+\sum_q g_T^{N,q}d_q\,,
\end{align}
with $N=p,n$.
We adopt the lattice inputs from Ref.~\cite{Liu:2024kqy}: $\theta_n=-1.48(34)\times10^{-3}\,e\,\fm$ and $\theta_p=+3.80(1.40)\times10^{-3}\,e\,\fm$, together with the proton tensor charges
\begin{align}
    g_T^{p,u}&=0.784(30) \, ,\nonumber \\[0.8mm]
    g_T^{p,d}&=-0.204(15) \, , \nonumber  \\[0.8mm]
    g_T^{p,s}&=-0.0027(16)\, ,
\end{align}
with neutron values obtained by interchanging $u$ and $d$. 
Combining the independent ILL and PSI neutron-EDM measurements~\cite{Pendlebury:2015lrz,Abel:2020pzs} gives $d_n^{\rm exp}=(-0.06\pm0.95)\times10^{-26}\, e\,\cm$, with experimental uncertainties added in quadrature.

For $^{199}$Hg, we include both the nucleon-EDM contribution and the contributions from the $\theta$-induced CP-odd pion--nucleon couplings. 
Following Ref.~\cite{Dekens:2022gha}, we use
\begin{align}
    \label{eq:Hg_fit}
    d_{\rm Hg}
    \simeq
    \kappa_{\rm at}
    \left[s_n d_n+s_p d_p+a_0\bar g_0+a_1\bar g_1\right],
\end{align}
where
$\kappa_{\rm at}=-(2.1\pm0.5)\times10^{-4}$, $s_n=1.9\pm0.1$, $s_p=0.20\pm0.06$, and $a_0= (0.13^{+0.50}_{-0.07})\,e\,\fm$, $a_1=(0.25^{+0.89}_{-0.63})\,e\,\fm$. 
For the pion--nucleon couplings, we use~\cite{deVries:2018mgf} $\bar g_0=-(14.7\pm2.3)\times10^{-3}\,\theta_{\rm eff},$ and $
\bar g_1=+(3.4\pm2.4)\times10^{-3}\,\theta_{\rm eff}$. Writing $\bar g_i=\lambda_i\theta_{\rm eff}$, we normalize the mercury observable to EDM units,
\begin{align}
    y
    \equiv
    \frac{d_{\rm Hg}}{\kappa_{\rm at}s_n}
    =
    d_n+r\,d_p+
    \frac{a_0\lambda_0+a_1\lambda_1}{s_n}\theta_{\rm eff}\,, 
\end{align}
with $r=s_p/s_n\simeq0.105$.
Using the central values of $\kappa_{\rm at}$ and $s_n$, the measurement of Ref.~\cite{Graner:2016ses} corresponds to $y_{\rm exp}=(-0.55\pm0.78)\times10^{-26}\,e\,\cm$.

We compare fits with fixed theoretical inputs to fits profiling over $\theta_n$, $\theta_p$, $a_0$, and $a_1$. Setting all $d_q=0$, the combined neutron--mercury fit yields $|\theta_{\rm eff}|\lesssim1.5\times10^{-10}$ ($\Delta\chi^2=4$) with this profiling prescription. We adopt this value for the benchmark (\textit{i}) discussed in the main text.

\begin{figure}[t]
    \centering   
    \vspace{5mm}\includegraphics[width=0.99\linewidth]{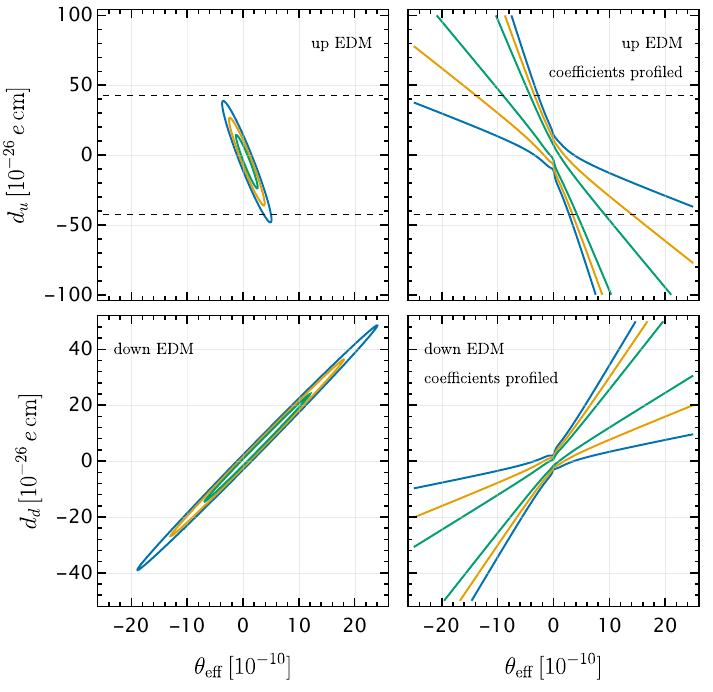}
    \caption{Confidence regions in the $(\theta_{\rm eff},\,d_q)$ plane for $q=u$ (upper panels) and $q=d$ (lower panels). The green, orange, and blue contours delimit the joint
    $1\sigma$, $2\sigma$, and $3\sigma$ regions.  In the left panels the  inputs $\theta_{n,p}$ and $a_{0,1}$ are held fixed at their central values. In the right panels they are profiled over. The dashed black lines delimit the band $|d_q|\le m_q/\Lambda_{\min}^{2}$ with $\Lambda_{\min}=10\,$TeV, large enough to evade complementary bounds from high energy searches~\cite{daSilvaAlmeida:2019cbr}.}
    \label{fig:theta_qEDM_bounds}
\end{figure}

We then allow either $d_u$ or $d_d$ to vary independently of $\theta_{\rm eff}$, setting the other quark EDMs to zero. Cancellations between the two CP-odd sources substantially enlarge the allowed region, as illustrated in Fig.~\ref{fig:theta_qEDM_bounds}. We consequently adopt $|\theta_{\rm eff}|=10^{-9}$ as a value for the benchmark (\textit{ii}) discussed in the main text.

We now consider the quark CEDM operators in Eq.~\eqref{eq:CEDM}, corresponding to benchmark~(\textit{iii}) in the main text. 
We allow $\tilde d_u$ and $\tilde d_d$ to vary
independently, setting all other higher-dimensional CP-odd sources to zero and assuming no explicit PQ breaking. 
Unlike in the preceding benchmarks, the residual angle is not an independent parameter but is generated by the CEDMs, as in Eq.~\eqref{eq:theta_CEDM}.

\begin{figure}[t]
    \centering
    \includegraphics[width=0.9\linewidth]{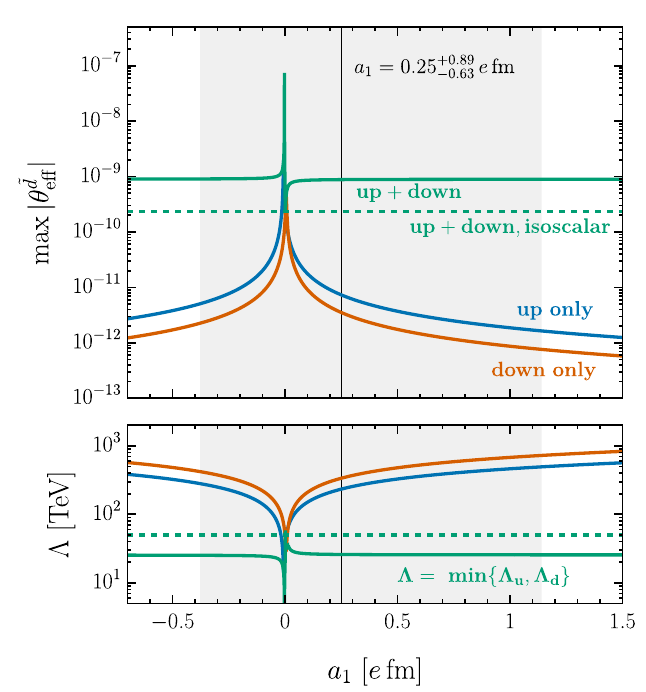}
    \caption{Upper panel: largest value of $\theta_ {\rm eff}^{\tilde{d}}$ compatible with the neutron and mercury EDMs as a function of the coefficient $a_1$. The curves correspond to a single up or down chromo-EDM, to the isoscalar combination $\tilde{d}_u=\tilde{d}_d$, and to two independent chromo-EDMs. The shaded band is the $1\sigma$ range $a_1=0.25^{+0.89}_{-0.63}\,e\,\text{fm}$~\cite{Dekens:2022gha}. Lower panel: new-physics scale $\Lambda_q=(m_q/|\tilde{d}_q|)^{1/2}$ of the chromo-EDMs that saturate the upper panel; for the two-flavour cases the smaller of $\Lambda_u$ and $\Lambda_d$ is shown.}
    \label{fig:maxTheta}
\end{figure}

To constrain these coefficients, we use the QCD sum-rule estimates for the nucleon EDMs in the presence of the PQ mechanism~\cite{Kaneta:2023wrl},
\begin{align}
    \label{eq:nucleon_EDM_CEDM}
    d_n&\simeq e\left(0.10\,\tilde d_u+0.20\,\tilde d_d\right),
    \nonumber\\
    d_p&\simeq-e\left(0.40\,\tilde d_u+0.049\,\tilde d_d\right),
\end{align}
with the Wilson coefficients evaluated at $\mu=2\,\GeV$. 
These expressions already incorporate the induced vacuum angle.

For mercury, we use Eq.~\eqref{eq:Hg_fit}, including both $\bar g_0$ and $\bar g_1$. 
Their CEDM-induced contributions contain both the indirect terms from vacuum alignment and the direct terms involving chromomagnetic nucleon matrix elements~\cite{Dekens:2022gha}. 
In the notation of our chiral Lagrangian, we adopt a common-ratio ansatz, $\tilde{c}_1/c_1=\tilde{c_5}/c_5$, and define $\xi_N\equiv(1-\tilde c_{1,5}/c_{1,5})$ expected to be $\sim\mathcal{O}(1)$. We thus find $ \bar g_0=-\xi_N\,C_0(\tilde d_u+\tilde d_d)$ and $\bar g_1=-\xi_N\,C_1 (\tilde d_u-\tilde d_d)$, where our hadronic inputs give $C_0\simeq2.10\,\GeV$ and $C_1\simeq39.6\,\GeV$.

In the benchmark, we fix $\xi_N=1$ and scan over the nuclear coefficient $a_1$. 
For each value of $a_1$, we identify the combination of $\tilde d_u$ and $\tilde d_d$ that maximizes $|\theta_{\rm eff}^{\tilde d}|$ while remaining compatible with the neutron and mercury EDM measurements at the two-standard-deviation level. 
The results are shown in Fig.~\ref{fig:maxTheta}. 
At the central value $a_1=0.25\,e\,\fm$, we obtain $|\theta_{\rm eff}^{\tilde d}|_{\max}\simeq8.8\times10^{-10}$.

\bibliography{nEDM_axions_andNon}

\end{document}